\documentclass[conference]{IEEEtran}
\IEEEoverridecommandlockouts
\usepackage{cite}
\usepackage{amsmath,amssymb,amsfonts}
\usepackage{algorithmic}
\usepackage{graphicx}
\usepackage{textcomp}
\usepackage{xcolor}
\def\BibTeX{{\rm B\kern-.05em{\sc i\kern-.025em b}\kern-.08em
    T\kern-.1667em\lower.7ex\hbox{E}\kern-.125emX}}
\begin{document}
\title{All Attention U-NET for Semantic Segmentation of Intracranial Hemorrhages In Head CT Images}
\author{\IEEEauthorblockN{Chia Shuo Chang, Tian Sheuan Chang}
\IEEEauthorblockA{\textit{Institute of Electronics} \\
\textit{National Yang Ming Chiao Tung University}\\
Hsinchu, Taiwan \\
johnson19961130@gmail.com, tschang@nycu.edu.tw}
\and
\IEEEauthorblockN{Jiun Lin Yan, Li Ko}
\IEEEauthorblockA{\textit{Dept. of Neurosurgery} \\
\textit{Chang Gung Memorial Hospital}\\
Taoyan, Taiwan\\
color\_genie@hotmail.com, lisa2017@cgmh.org.tw}
}
\maketitle             
%
\begin{abstract}
Intracranial hemorrhages in head CT scans serve as a first line tool to help specialists diagnose different types. However, their types have diverse shapes in the same type but similar confusing shape, size and location between types. To solve this problem, this paper proposes an all attention U-Net. It uses channel attentions in the U-Net encoder side to enhance class specific feature extraction, and space and channel attentions in the U-Net decoder side for more accurate shape extraction and type classification. The simulation results show up to a 31.8\% improvement compared to baseline, ResNet50 + U-Net, and better performance than in cases with limited attention. 

\end{abstract}

\section{Introduction}

Head CT image scan is a widely accepted first-line tool in the emergency room for head injuries, stroke, or any other intracranial lesion because of its short acquisition time. To identify the individual lesion type and area is important for diagnosis but also time-consuming, especially for small and urgent diseases. A promising and well performed approach in recent years is to use deep learning for this medical image analysis, which has been used in  identification of diabetic retinopathy\cite{gulshan2016development}, classification of skin lesions\cite{esteva2017dermatologist}, abnormality detection in chest CT\cite{anthimopoulos2016lung}, X-ray images\cite{wang2017chestx}, and head CT\cite{gao2017classification}. For our target problem, intracranial hemorrhages in head CT scans, semantic segmentation can help locate and quantify these disease patterns. However, it is not easy to distinguish seven types of intracranial hemorrhages in head CT scans because the same type has different shapes and different types have similar shapes, sizes, and locations, which poses a challenge to the deep learning network.

Semantic segmentation of intracranial hemorrhages in head CT scans also faces the challenge of high resolution input (up to 512x512), which is not feasible to resize the image for training and testing due to some extremely small size lesions. Furthermore, this also prevents the use of some advanced models that require large memory during training\cite{wang2020deep} \cite{chen2017rethinking}\cite{chen2018encoder} and thus makes the batch size too small or even impossible to train. Among the deep learning models, U-Net \cite{ronneberger2015u} has been widely used in medical image analysis due to its smaller training cost and outstanding performance even with few data as in the medical image case\cite{christ2017automatic}. However, the original U-Net architecture is made up of stacked convolutions without enhancing the dependence of the difference channel and layer information. This makes it difficult to extract global features and generate correct shapes for medical image analysis.

Various enhanced versions of U-Net have been proposed for better performance. One approach is to adopt the network from the latest classification network as the backbone of U-Net architecture\cite{he2016deep}\cite{pleiss2017memory}\cite{szegedy2016rethinking}. Another way is to use different path aggregation methods to combine different levels of features\cite{zhou2018unet++}. However, these aggregation methods, in which the aggregation cells are connected through a series of nested and dense skip pathways, consume too much computation and memory. Besides, for the target problem, their performance is still limited as shown in our experimental results. Another approach is to use the attention mechanism that can increase the receptive field and improve the long-range dependencies in the model without adding too much computation and parameters. These attention networks include attentions either on space\cite{zhao2018psanet},  channel\cite{hu2018squeeze}, or group\cite{li2019selective} dimensions, which has been integrated into the U-Net decoder side\cite{oktay2018attention}\cite{abraham2019novel}\cite{sinha2020multi}. However, the channel maps of the input feature map on the decoder side are still disrupted, which demands further improvement, especially for the lesion dataset with high variability. In addition, none has considered the issue of attention on the encoder side.

To solve the above issue, we propose the all attention U-Net, which improves the U-Net encoder with the channel attention and U-Net decoder with space and channel attention to resolve the conflicts of shape, size and location. The simulation result shows that the proposed network can predict a variety of intracranial  hemorrhage but still with small model size, which is better than a baseline ResNet50 + U-Net and U-Net with only one kind of attention.

\begin{figure*}[!htb]
\centering
\includegraphics[width=1\textwidth]{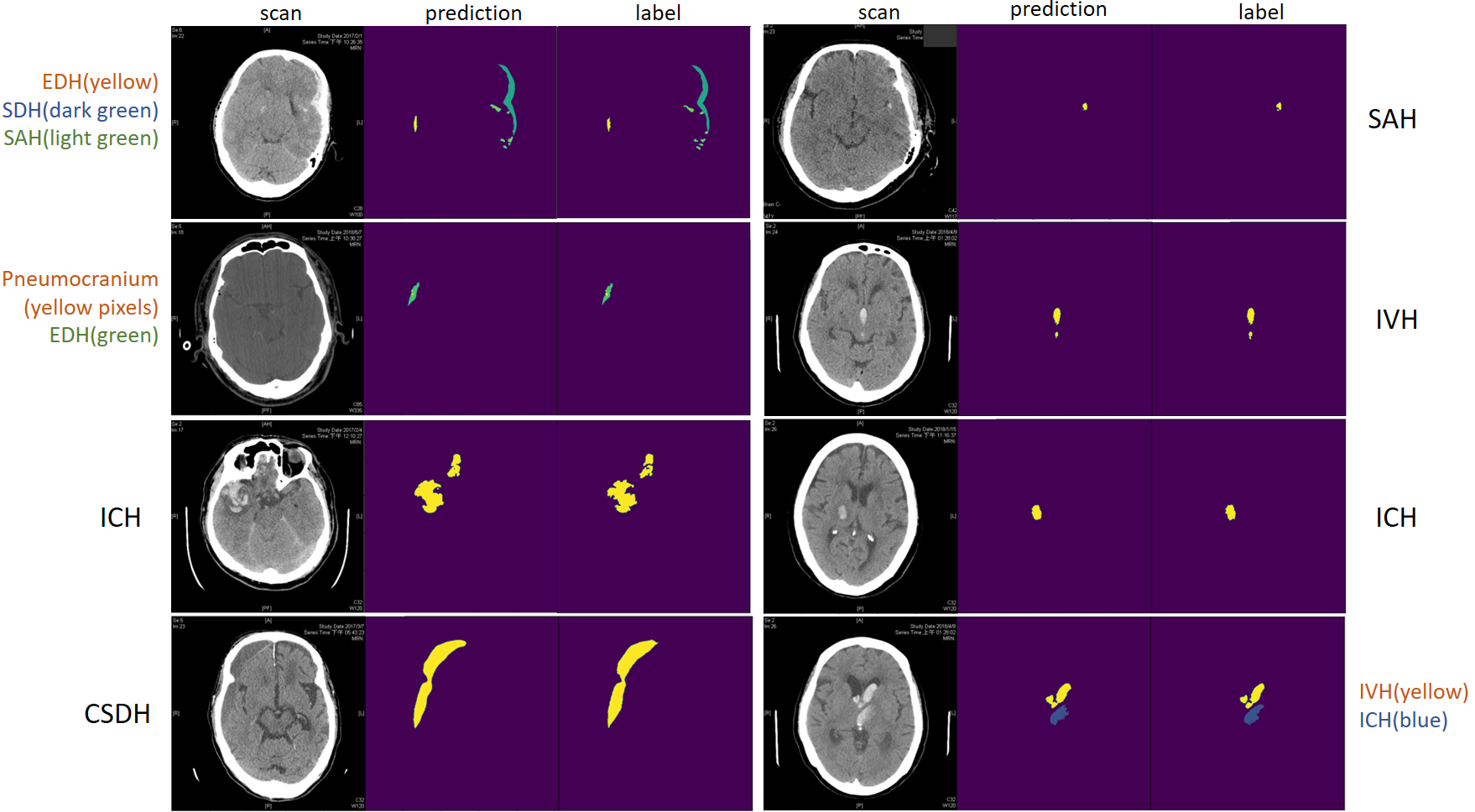}
\caption{Example of different intracranial hemorrhages, where the prediction is the result based on our model.}
\label{Fig.total}
\end{figure*}

\section{DATASET}

The dataset consisted of a total of 51 cases, approximately 2,048 head CT scans from Chang Gung Memorial Hospital, Taiwan. These CT scan images have been labeled by a specialist physician. Data usage was approved by the local research ethics committee. The slice numbers vary from 30 to 50 per case. These cases contain 7 kinds of intracranial hemorrhages, namely intracerebral
hemorrhage (ICH), acute subdural hematoma (SDH), subarachnoid hemorrhage (SAH), extradural hematoma (EDH), chronic subdural
hematoma (CSDH), Pneumocranium, intraventricular hemorrhage (IVH).

\subsection{Data Imbalance And High Resolution Images}

\begin{figure}[!htb]
\centering
\includegraphics[width=1.0\linewidth]{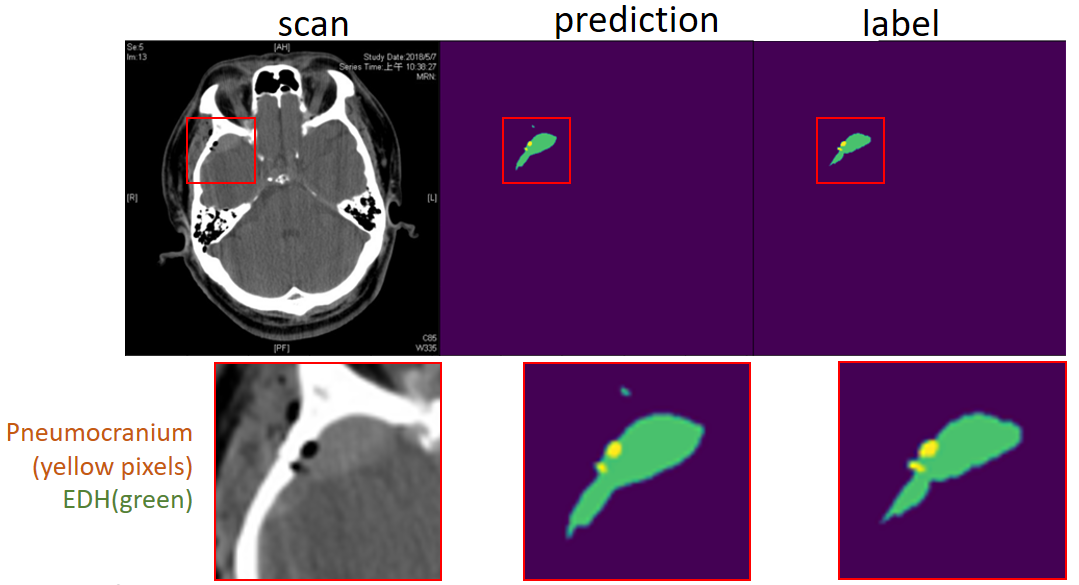}
\caption{Examples of the Pneumocranium and EDH. The area of Pneumocranium is very small in all our cases.}
\label{Fig.big}
\end{figure}

As in other medical image dataset, our dataset is also small because of the difficulty of ground truth labeling. Besides, many scans in a case show no lesions(about 46.55\%) and the area of the lesion is small in most of the scans as shown in Fig.~\ref{Fig.total}. This data imbalance will cause training difficulties. Besides, the size of CT scan images is large, 512 x 512. High resolution input will consume too much GPU memory and take long training time. However, these images cannot be resized because the very small area in some types of lesion (fewer than 20 pixels as in Fig.~\ref{Fig.big}) will be removed from the resized image.

\begin{figure}[!htb]
\centering
\includegraphics[width=1.0\linewidth]{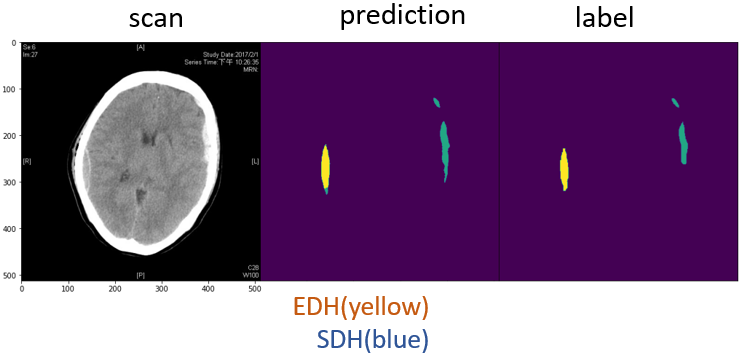}
\caption{Features of EDH and SDH are very similar, but with different shapes. EDH is 
fusiform and SDH is crescent.}
\label{Fig.similar}
\end{figure}

\subsection{High Variety of Lesions}

Fig.~\ref{Fig.total} shows examples of different intracranial hemorrhages. One image could contain one to several types of lesions. However, if similar lesions appear at the same time, it will be difficult to distinguish, even by specialists. For example, EDH and SDH are similar in their textures and locations (Fig.~\ref{Fig.similar}), and the only difference is their shapes. Other confusing lesions are SAH, IVH and ICH, which are similar in shape and size for some cases. Furthermore, even in the same type of lesion, its shape and size also have many variations, such as the ICH shown in Fig.~\ref{Fig.total}. CSDH can be decided by its texture feature, but not all this type of feature is CSDH. The area of Pneumocranium is very small as shown Fig.~\ref{Fig.big}. Its total number of pixels is only from 15 to 86 pixels in each scan. The specialist always needs to magnify the CT images to observe this lesion, which is very laborious and time-consuming. These all pose challenges to the network design.

\section{NETWORK ARCHITECTURE}

\begin{figure}[!htb]
\centering
\includegraphics[width=1.0\linewidth]{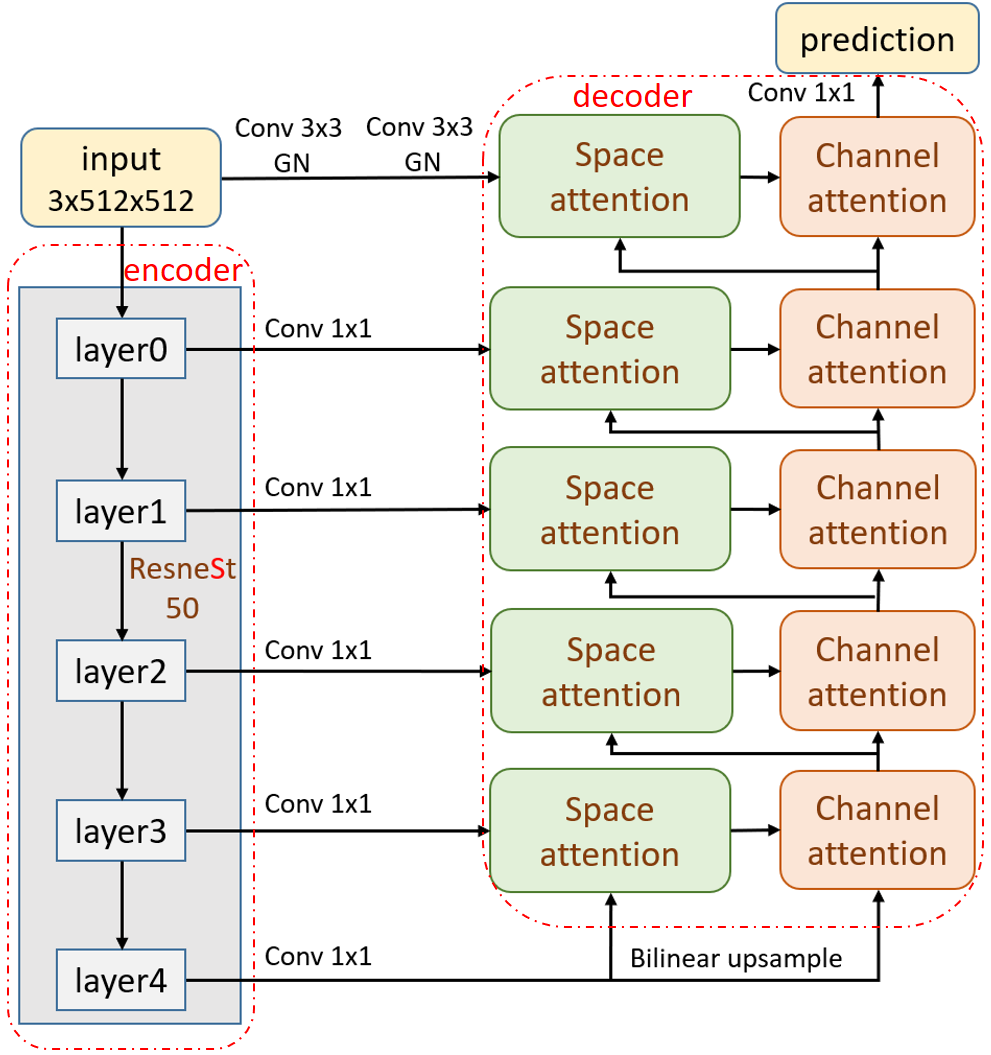}
\caption{All Attention U-Net architecture}
\label{Fig.network}
\end{figure}

Fig.~\ref{Fig.network} shows our proposed network based on the famous U-Net \cite{ronneberger2015u} that contains encoder and decoder paths with direct skip connections between encoder and decoder paths. The encoder side adopts the state-of-the-art classification network, ResNeSt50\cite{zhang2020resnest}, as the backbone, since it combines channel attention and ResNet to enrich feature extraction. It alters the residual block by splitting the channels into groups and performing channel attention on them to highlight the important channel groups and suppress the unnecessary ones.
In this paper, to further selectively enhance the desired features, the extracted feature maps of different resolutions are first selectively enhanced by space attention for shape and then aggregrated by channel attention with upsampling in the decoder paths, as illustrated in Fig.~\ref{Fig.network}. In addition, unlike direct skip connections in the original U-Net, in this paper, the skip connections between encoder and decoder consist of 1x1 convolution and ReLU to reduce the number of channels for lower complexity. In all these skip connections, to retain precise location information, the lowest level features are from the input images directly after two 3x3 convolutions with ReLU layers. For the decoder path, this paper adds the channel attention blocks to repair the disrupted channel maps for better class and shape prediction.  


\begin{table*}[htb]
\centering
\caption{Dice score coefficient for segmentation results on 7 types of intracranial hemorrhage. The network name like ResNet 50 + U-Net means U-Net with ResNet50 as the encoder. ('RSU' : ResNeSt50 + U-Net; 'S': space attention; 'C': channel attention)}
\begin{tabular}{llllllll}

\hline
Network                     & ICH            & SDH           & SAH            & EDH             & CSDH           & Pneumocranium  & IVH             \\ \hline
ResNet50 + U-Net                              & 0.9015        & 0.534        & 0.316          & 0.498          & 0.845        & 0.7            & 0.794           \\
DenseNet121 + U-Net                          & \textbf{0.933} & 0.731         & 0.375         & 0.607           & 0.864        & \textbf{0.76}  & \textbf{0.87}   \\
UNet++\cite{zhou2018unet++}                                       & 0.846         & 0.58          & 0.39          & 0.436          & 0.88 & 0.654           & 0.81          \\
Attention U-Net\cite{oktay2018attention} & 0.9135 & 0.74 & 0.465 & 0.63 & 0.857 & 0.684 & 0.793 \\
HRNet\cite{wang2020deep}                                      & 0.812         & 0.42          & 0.251          & 0.446          & 0.872 & 0.554           & 0.45          \\ \hline
RSU(ours)                             & 0.914          & \textbf{0.8}  & 0.45           & 0.74            & 0.877          & 0.68           & 0.82            \\
RSU + S(ours)               & 0.9318         & 0.78          & 0.39           & 0.63            & \textbf{0.905} & 0.634          & 0.792         \\
RSU + C(ours)             & \textbf{0.932} & 0.777         & \textbf{0.53}  & \textbf{0.779}  & 0.843          & \textbf{0.743} & \textbf{0.868} \\
RSU + SC(ours) & 0.924          & \textbf{0.82} & \textbf{0.567} & \textbf{0.816} & \textbf{0.906} & \textit{0.71}  & \textit{0.858}  \\ \hline
\end{tabular}
\label{Table.result}
\end{table*}
\subsection{Decoder Side - Channel Attention}


Channel maps can be regarded as class-specific responses. However, the channel maps are disrupted after a series of path aggregation and transformation. To solve this problem, we add channel attention \cite{hu2018squeeze} into the decoder side to make contextual representation more class specific. 
We concatenate the output from the previous layer and space attention block, upsample the result by two and then apply channel attention module \cite{hu2018squeeze} with global average pooling, two fully connected layers (FC) and sigmoid function. In this block, we use Group Normalization (GN) instead of Batch Normalization (BN) to better fit into GPU memory size and also get better performance. 

\subsection{Path Aggregation With Space Attention}

\begin{figure}[!htb]
\centering
\includegraphics[width=1.0\linewidth]{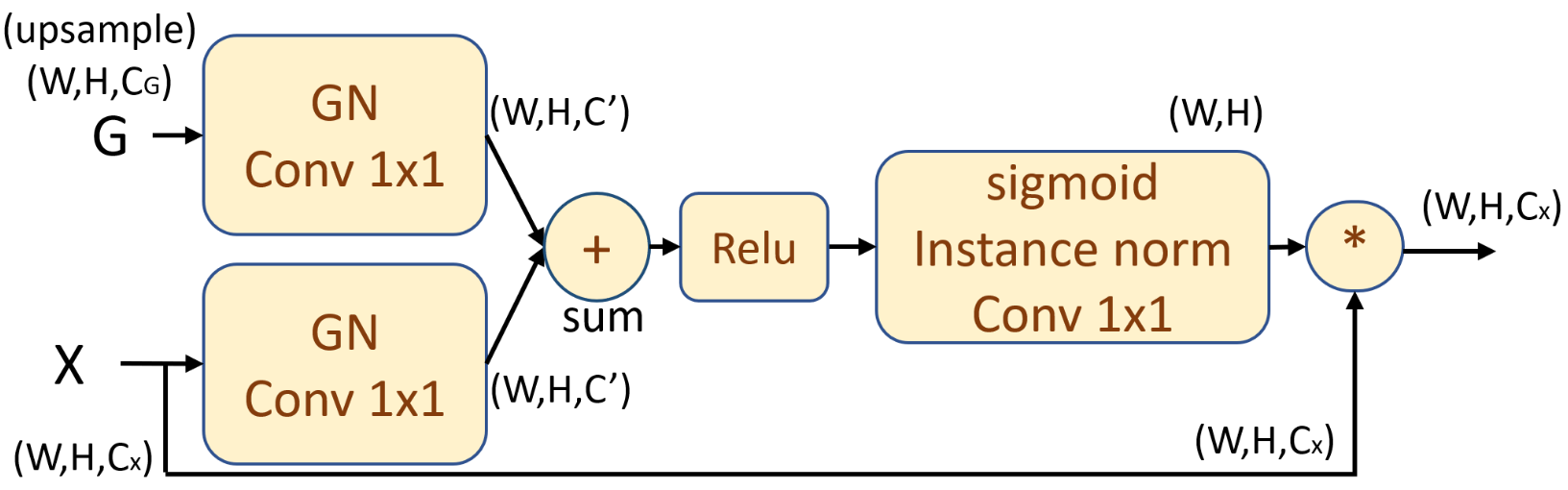}
\caption{Space Attention block in path aggregation part}
\label{Fig.self_att}
\end{figure}

For intracranial hemorrhages in head CT scans, location information is one important feature because some kinds of lesions only occur in certain areas. Certain lesions have very different shape and size that makes model hard to predict correctly. Besides, some lesions are very small that its spatial details are easily lost with cascaded convolutions and transformation.

To solve these problems, this paper adopts space attention \cite{oktay2018attention} as shown in Fig.~\ref{Fig.self_att} to automatically learn to focus on the specific lesion area and thus eliminate the necessity of post-processing. This block also aggregates the feature map from different layers by merging adjacent different resolution features from channel attention and applying space attention on it. Here, we adopt GN and instance normalization (IN) instead of BN for better input specific results.

\section{RESULT}
\subsection{Experimental Setting}
The model uses three consecutive CT images as input to predict the center image more accurately by combining neighboring information. The input is further augmented by many auto-augmentation techniques, like random rotation, crop, flipping, altering contrast, brightness, and saturation, to avoid overfitting caused by too few data. The model is trained on four NVIDIA RTX-2080Ti GPUs with the AdamW optimizer\cite{loshchilov2017decoupled}, cyclic learning rates\cite{smith2017cyclical}, and mixed precision. Furthermore, the loss function is class-weighted focal loss \cite{lin2017focal} to alleviate the data imbalance problem.  We use the Dice score coefficient as our evaluation metrics and split the dataset into training(78\%), validation(7.5\%) and testing set(14.5\%). It is worth noting that, to avoid overestimation of accuracy, we partition the dataset based on different patient cases instead of randomly selecting slice images, because the distribution of the same patient case is similar.

\begin{figure}[htbp]
  \centering

  \includegraphics[height=!,width=1.0\linewidth,keepaspectratio=true]
  {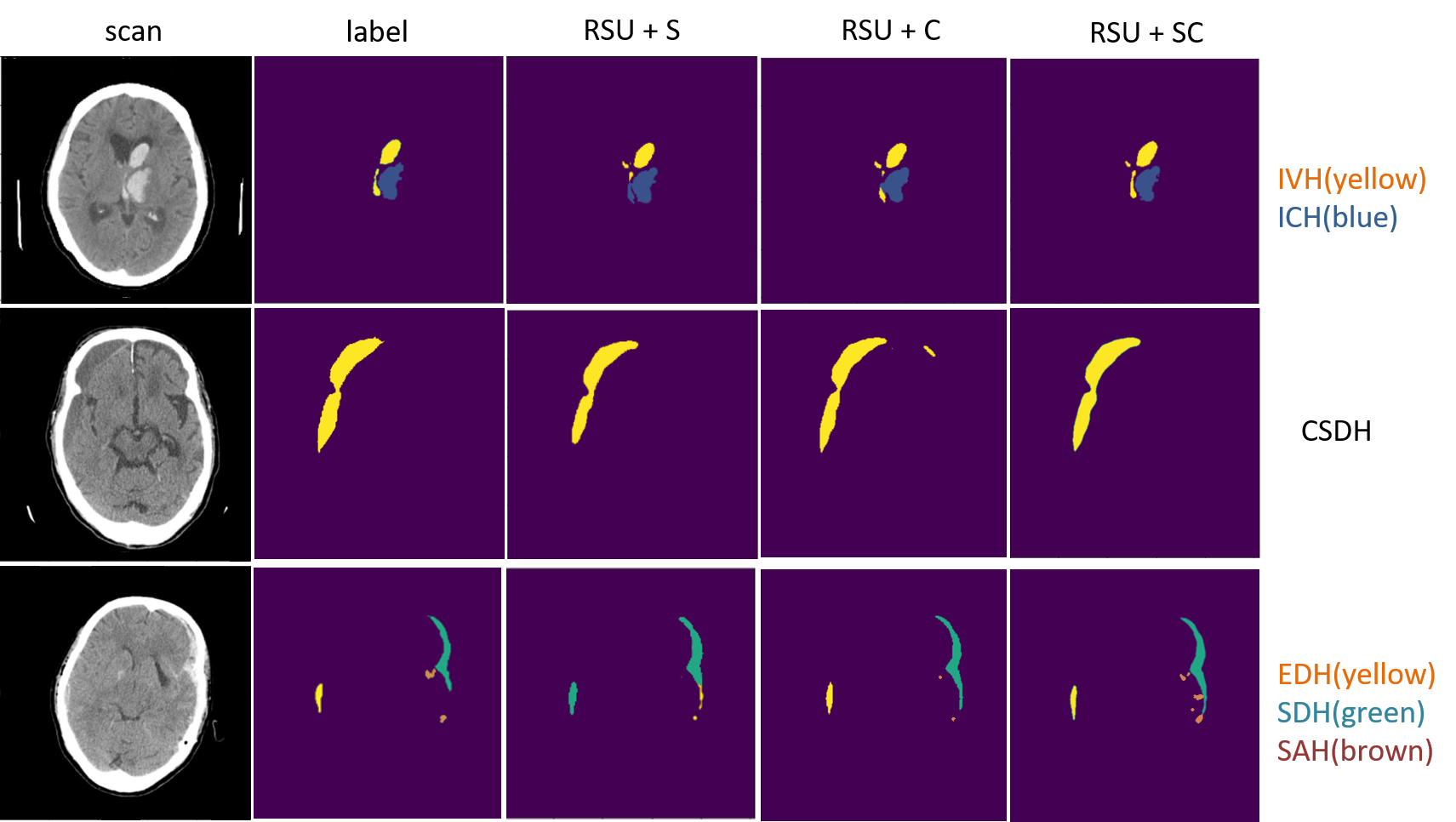}
  \caption{Prediction results of our network.( 'RSU' : ResNeSt50 + U-Net ;  'S' : space attention ; 'C' : channel attention)}
  \label{fig.miccai_result}
\end{figure}

\subsection{Result}
Table.~\ref{Table.result} shows the evaluation results. The accuracy of the proposed network (RSU + SC) is the best in most of the lesion cases, and close to the best in other lesion cases, which shows our strong network generalization capability for these challenging types of lesions. In comparison, previous networks as shown in Table.~\ref{Table.result} have difficulties learning well about some specific lesions like SDH, EDH and SAH due to their highly similar features. The prediction results are shown in Fig. ~\ref{Fig.total} and \ref{fig.miccai_result}.

For the attention scheme, space attention and channel attention have their own advantages. The use of space attention in the path aggregation part can  improve CSDH and ICH accuracy significantly because they usually occur in certain specific areas. The space attention helps extract the location information and makes the shape more accurate. Channel attention in the decoder part improves detection for most cases because it reduces the chance of false identification and makes the prediction shape more accurate.


Finally, the full version of the proposed network (RSU + SC) combines the advantages of spatial and channel attention. It solves the problem of disrupted channel maps caused by aggregation of paths. Compared to the baseline, ResNet50 + U-Net, the improvement in the Dice coefficient of each lesion in our proposed model is 2.25\%, 28.6\%, 25.1\%, 31.8\%, 6.1\%, 1\% for ICH, SDH, SAH, EDH, CSDH, pneumocranium, IVH.

\section{CONCLUSION}
This paper proposes an all-attention U-Net to solve the challenging segmentation problem of intracranial hemorrhages on CT scans. The proposed network uses channel attention to highlight class-specific features and space attention to highlight disease related features. The attention scheme is integrated into both sides of U-Net to get the maximum benefit without increasing too many parameters. The evaluation result shows an improvement of up to 31.8\% compared to baseline and outperforms other segmentation networks.

\section{Acknowledgments}
This study is partially sponsored by the Chang Gung Memorial Hospital Research Project (CMRPG2H0371) and is supported in part by the Ministry of Science and Technology of Taiwan, under Grant 109-2634-F-009-022. The authors also thank the National Center for High-performance Computing (NCHC) for providing computational and storage resources.

%
%
%
%

\bibliographystyle{IEEEtran}
\bibliography{refs.bib}
\end{document}